\documentclass[vecphys]{svmult}

\usepackage{makeidx}          
\usepackage{graphicx}         
\usepackage{multicol}         
\usepackage[bottom]{footmisc}
\usepackage{amsmath,amssymb}

\newcommand{\m}{\widetilde m_u}
\newcommand{\M}{\widetilde M_u}
\newcommand{\md}{\widetilde m_d}
\newcommand{\Md}{\widetilde M_d}
\newcommand{\pslash}{\!\not\! p}

\makeindex             

\newcommand{\eps}{\epsilon}

\begin{document}

\title*{A holographic composite Higgs model}

\author{Roberto Contino
}

\institute{Dipartimento di Fisica, Universit\`a di Roma ``La Sapienza'' and \\
INFN Sezione di Roma, Piazzale Aldo Moro 2, I-00185 Roma, Italy
\texttt{roberto.contino@roma1.infn.it}
}

\maketitle

Composite Higgs models~\cite{GK} represent an attractive variation of the 
Technicolor paradigm~\cite{TC}. 
In these theories the Standard Model (SM) Higgs doublet is the bound state 
of a strongly interacting sector with flavor symmetry $G$.
It forms at a scale $f_\pi$, the analog of the 
QCD pion decay constant, as the Goldstone boson associated with the
dynamical breaking of the global symmetry $G$.
The couplings of the SM matter and gauge fields to the strong sector 
break $G$ explicitly,
and induce a one-loop Higgs potential that triggers the electroweak
symmetry breaking (EWSB) at the scale $v \leq f_\pi$.
In the limit of a large separation $\eps = v/f_\pi \ll 1$, all the massive bound states
of the strong sector decouple, and one is left with the low-energy spectrum of the Standard Model.
This means that all corrections to the electroweak precision observables constrained by
LEP and SLD experiments will be suppressed by powers of $\eps$.

In this talk I will discuss an interesting realization of this old idea
which has been recently found in the
framework of  extra-dimensional warped models~\cite{Contino:2003ve,Agashe:2004rs} 
defined on AdS$_5$ spacetime with two boundaries~\cite{Randall:1999ee}.
The main virtue of these models is  their calculability as effective field theories:
various physical quantities, like for example the Higgs potential
and the  electroweak precision observables, can be computed in 5D in a perturbative expansion.
A minimal model was introduced in~\cite{Agashe:2004rs}, where 
a bulk SO(5)$\times$U(1)$_X\times$SU(3)$_c$ gauge symmetry is
reduced to the SM group G$_{\rm SM}$=SU(2)$_L\times$U(1)$_Y\times$SU(3)$_c$
on the UV boundary and to SO(4)$\times$U(1)$_X\times$SU(3)$_c$ on the IR boundary.
Hypercharge is defined as $Y=T^3_{R}+X$, where SO(4)$\sim$SU(2)$_L\times$SU(2)$_R$.
According to the AdS/CFT correspondence~\cite{AdSCFT}, such 5-dimensional scenario 
is equivalent to a
4D composite Higgs theory where the SO(5) flavor symmetry of the strong sector is
spontaneously broken to SO(4) in the infrared.
This delivers 4 Goldstone bosons that transform as a \textbf{4} of SO(4)
(a real bidoublet of SU(2)$_L\times$SU(2)$_R$) and are identified with the Higgs doublet.
They correspond, in the 5D theory, to the SO(5)/SO(4) degrees of freedom of the fifth component
$A_5$ of the bulk gauge field.

In~\cite{Agashe:2004rs} the SM fermions were embedded in spinorial representations of the
bulk SO(5) gauge symmetry. This choice leads to a generically large modification of the 
$Zb\bar b$ coupling, which in turn rules out a large portion of the 
parameter space~\cite{Agashe:2005dk}.
A more natural model, however, can be simply obtained if the bulk SO(5) symmetry is reduced
to O(4) on the IR brane, instead of SO(4), and by embedding the SM fields in fundamental
(\textbf{5}) or antisymmetric (\textbf{10}) representations of SO(5). 
In this case, indeed, a subgroup of the custodial symmetry O(3) that protects the
electroweak $\rho$ parameters from large corrections, also protects the 
$Zb\bar b$ coupling~\cite{Agashe:2006at}. 
A possible choice of bulk fields and boundary conditions is the following~\cite{next}:
\begin{equation} \label{reps}
\begin{split}
\xi_{q_1} (\mathbf{5}_{+2/3}) &= 
 \begin{bmatrix}
 q'_{1L}(-+) & q'_{1R}(+-) \\
 q_{1L}(++)  & q_{1R}(--)  \\
 s^u_L(--)   & s^u_R(++)
 \end{bmatrix}
 \, , \\[0.2cm]
\xi_{q_2} (\mathbf{5}_{-1/3}) &= 
 \begin{bmatrix}
 q_{2L}(++)   & q_{2R}(--) \\
 q'_{2L}(-+)  & q'_{2R}(+-)  \\
 s^d_L(--)    & s^d_R(++)
 \end{bmatrix}\, ,
\end{split}\, \quad
\begin{split}
\xi_u (\mathbf{5}_{+2/3}) &= 
 \begin{bmatrix}
 q'^{u}_L(+-) & q'^{u}_R(-+) \\
 q^{u}_L(+-)  & q^{u}_R(-+)  \\
 u_L(-+)      & u_R(+-)
 \end{bmatrix} 
  \, , \\[0.2cm]
\xi_d (\mathbf{5}_{-1/3}) &= 
 \begin{bmatrix}
 q^{d}_L(+-)  & q^{d}_R(-+) \\
 q'^{d}_L(+-) & q'^{d}_R(-+)  \\
 d_L(-+)  & d_R(+-)
 \end{bmatrix} \, .
\end{split}
\end{equation}
Chiralities under the 4D Lorentz group have been denoted with $L$, $R$, and 
$(\pm,\pm)$ is a shorthand notation to denote Neumann $(+)$ or Dirichlet $(-)$
boundary conditions.
The fields $\xi_{q_1}$,  $\xi_{u}$ ($\xi_{q_2}$,  $\xi_{d}$) transform as \textbf{5}$_{2/3}$
(\textbf{5}$_{-1/3}$) of SO(5)$\times$U(1)$_X$, and their
zero modes are identified with a full generation of SM quarks.
In particular, only a linear combination of $q_{1L}$ and $q_{2L}$
has Neumann boundary condition on the UV brane -- its zero mode
is identified with the SM quark doublet $q_L$ -- the other combination
being Dirichlet.~\footnote{The same can be obtained by starting with both
$q_{1L}$ and $q_{2L}$ having Neumann UV boundary conditions and by 
adding a  mass mixing term on the UV brane.} 
A fundamental of SO(5) decomposes as 
$\mathbf{5} = \mathbf{4} + \mathbf{1}  = \mathbf{(2,2)} + \mathbf{(1,1)}$
under SO(4)$\sim$SU(2)$_L\times$SU(2)$_R$, and we have defined the
$(q,q')$ fields to transform as $\mathbf{(2,2)}$'s, while $s^u$, $s^d$, $u$, $d$ transform as singlets.
A similar 5D embedding also works for the SM leptons, although with different U(1)$_X$ charges.
Localized on the IR brane, we consider the most general O(4)-invariant set of mass terms:
\begin{equation*}
 \m \left( \bar q_{1L} q^{u}_R + \bar q'_{1L} q'^{u}_R \right) + \M \,\bar s^u_R u_L +
 \md \left( \bar q_{2L} q^{d}_R + \bar q'_{2L} q'^{d}_R \right) + \Md\, \bar s^d_R d_L + h.c. 
\end{equation*}

The most general low-energy Lagrangian for the SM fermions that follows from the
embedding (\ref{reps}) is, in momentum-space and at the quadratic order~\cite{next}:
\begin{equation} \label{Lfermions}
\begin{split}
{\cal L}_{eff}= 
&\bar q_L \pslash \left[ \Pi_0^q 
 + \frac{s^2}{2} \left( \Pi_1^{q1}\, \hat H^c \hat H^{c\dagger} 
 +  \Pi_1^{q2}\, \hat H \hat H^\dagger \right) \right] q_L
 +\bar u_R \pslash \left( \Pi_0^u + c^2 \Pi_1^u \right) u_R \\
&+\bar d_R \pslash \left( \Pi_0^d + c^2 \Pi_1^d \right) d_R 
 + \frac{sc}{\sqrt{2}} M_1^u \,\bar q_L \hat H^c u_R 
 + \frac{sc}{\sqrt{2}} M_1^d \,\bar q_L \hat H d_R + h.c.
\end{split}
\end{equation}
Here $c=\cos(h/f_\pi)$, $s=\sin(h/f_\pi)$,  and $h = \sqrt{\left( h^a\right)^2}$,
where $h^{a}$ denote the four real components of the Higgs field:
\begin{equation}
\hat H = \frac{1}{h} 
 \begin{bmatrix} \hat h_1 - i \hat h_2 \\ \hat h_3 - i \hat h_4 \end{bmatrix}\, , \quad
\hat H^c = \frac{1}{h}
 \begin{bmatrix} -(\hat h_3 + i \hat h_4) \\ \hat h_1 + i \hat h_2 \end{bmatrix}\, .
\end{equation}
The form factors $\Pi(p^2)$, $M(p^2)$ can be computed in terms of 5D propagators
using the holographic approach of~\cite{Agashe:2004rs}.
The analog low-energy Lagrangian for the SM gauge fields can be found in \cite{Agashe:2004rs};
its form implies the following relations:
\begin{equation}
v \equiv \eps f_\pi = f_\pi \sin\frac{\langle h\rangle}{f_\pi} = 246\; \text{GeV}\, , \qquad  
 f_\pi = \frac{2}{\sqrt{g_5^2 k}} \frac{1}{L_1}\, ,
\end{equation}
where $g_5$ is the SO(5) gauge coupling in the bulk, $k$ is the AdS$_5$ curvature,
and $L_1$ is the position of the IR brane which sets the mass scale of the 
new particles ($1/L_1 \sim$ TeV).

The Higgs potential is generated at  one loop  from the virtual exchange of SM fields.
The largest contribution comes from those fields that couple more strongly to the Higgs, 
namely $t_R$, $t_L$ and $b_L$:
\begin{equation} \label{potential}
\begin{split}
V(h) =& (-6) \int\!\frac{d^4p}{(2\pi)^4}\, \Big\{ 
 \log \Big( \Pi_0^q + \frac{s^2}{2} \Pi_1^{q2} \Big) \\
 & + \log\Big[ p^2 \left( \Pi_0^u + c^2 \Pi_1^u \right) \Big( \Pi_0^q + \frac{s^2}{2} \Pi_1^{q1} \Big)  
              -\frac{s^2 c^2}{2}  (M_1^u)^2 \Big] \Big\}\, ,
\end{split}
\end{equation}
where the form factors 
refer to $q_L=(t_L,b_L)$ and $t_R$.
Since $\Pi_1$ and $M_1$ drop exponentially for $pL_1 \gg 1$, 
the logarithm in Eq.~(\ref{potential}) can be expanded and the potential is well approximated by:
\begin{equation}
V(h) \simeq \alpha\, \sin^2\frac{h}{f_\pi} + \beta \sin^4\frac{h}{f_\pi}\, ,
\end{equation}
where $\alpha$ and $\beta$ are integral functions of the form factors.
For $\alpha < 0$ and $2 \beta>|\alpha|$ the electroweak symmetry is broken, and
the minimum is at
\begin{equation}
\sin\frac{\langle h\rangle}{f_\pi} = \eps = \sqrt{\frac{-\alpha}{2\beta}}\, .
\end{equation}
If $\alpha < 0$ and $2\beta \leq |\alpha|$, on the other hand, $\cos\langle h\rangle/f_\pi=0$
and the EWSB is maximal: $\eps=1$. In this limit an O(4) chiral symmetry
is restored, and all the Yukawa couplings (hence the fermion masses) vanish, as one can 
explicitly see from the effective Lagrangian (\ref{Lfermions}).
The model is thus realistic only for $0<\eps<1$, which can be obtained for natural values of the
5D input parameters~\cite{next}.
Determining how much of this region is excluded by the electroweak precision constraints 
gives a measure of how natural is the model in reproducing the EWSB.
The strongest bound comes from the Peskin-Takeuchi $S$ parameter:
\begin{equation} \label{Spar}
S = \frac{6\pi}{g_5^2 k}\, \eps^2\, .
\end{equation}
A rough estimate using Eq.~(\ref{Spar}) suggests that imposing $S\leq 0.3$ excludes
$\sim 50\%$ ($\sim 75\%$) of the $0<\eps<1$ region for
$1/N = 1/5$ ($1/N = 1/10$), where $1/N \equiv (g_5^2k/16\pi^2)$ is the
5D expansion parameter. 
A detailed numerical analysis gives similar results~\cite{next}.
This shows that a sizable portion of the parameter space is still allowed,
and that no large fine-tuning is required in this model to pass all 
electroweak precision tests.
Moreover, the model predicts a light physical Higgs: 
$100\; \text{GeV} \lesssim m_\text{Higgs} \lesssim 150\; \text{GeV}$.
This is possible, and the ``little hierarchy'' puzzle is resolved, since
the spectrum of the new vectorial states (which enter the oblique precision observables) 
is predicted to be heavier than the fermionic resonances 
(which are responsible for cutting off the SM top loop).
In fact, the most important prediction  of the model 
is that, due to the heaviness of the top quark,
at least one among its Kaluza-Klein partners is relatively light, with a mass 
of order $500-1500$ GeV. These new fermions transform as $SU(2)_L$ doublets with hypercharge
$Y=1/6$ or $7/6$, and singlets with $Y=2/3$.
They will be both singly and pair produced at the LHC, 
decaying to final states populated mostly by tops, bottoms 
and SM gauge and Higgs bosons.
Studying their phenomenology at the LHC will be certainly 
exciting and challenging at the same time.

\acknowledgement
I would like to thank K.~Agashe, A.~Pomarol and R.~Sundrum for the numerous
inspiring discussions and an always stimulating collaboration.
I also thank the organizers of IFAE 2006 for inviting me to this
interesting conference.

%
%


\printindex
\end{document}